\documentclass{PoS}

\title{Update of the Unitarity Triangle Analysis}

\ShortTitle{Update of the UTA}

\author{A.~J.~Bevan, M.~Bona\\
       Queen Mary, University of London, E1 4NS, United Kingdom}
\author{M.~Ciuchini\\
        INFN,  Sezione di Roma Tre, I-00146 Roma, Italy}
\author{D.~Derkach, A.~Stocchi\\
       Laboratoire de l'Acc\'el\'erateur Lin\'eaire, IN2P3-CNRS et
       Universit\'e de Paris-Sud, BP 34, 
       F-91898 Orsay Cedex, France}
\author{E.~Franco, L.~Silvestrini\\
        INFN, Sezione di Roma, I-00185 Roma, Italy}
\author{V.~Lubicz, \speaker{Cecilia Tarantino}\\
        Dipartimento di Fisica, Universit{\`a} Roma Tre, and INFN, I-00146 Roma, Italy} 
\author{G.~Martinelli\\
        Dipartimento di Fisica, Universit\`a di Roma ``La
        Sapienza'', and INFN, I-00185 Roma, Italy} 
\author{F.~Parodi, C.~Schiavi\\
       Dipartimento di Fisica, Universit\`a di Genova and INFN, I-16146
       Genova, Italy} 
\author{M.~Pierini\\
       CERN, CH-1211 Geneva 23, Switzerland}
\author{V.~Sordini\\
       IPNL-IN2P3 Lyon, France}
\author{V.~Vagnoni\\
        INFN, Sezione di Bologna,  I-40126 Bologna, Italy}

\abstract{We present the status of the Unitarity Triangle Analysis (UTA), within the Standard Model (SM) and beyond, with experimental and theoretical inputs updated for the ICHEP 2010 conference .
Within the SM, we find that the general consistency among all the constraints leaves space only to some tension (between the UTA prediction and the experimental measurement) in BR$(B \to \tau \nu)$, $\sin 2 \beta$ and $\varepsilon_K$.
In the UTA beyond the SM, we allow for New Physics (NP) effects in $\Delta F=2$ processes.
The hint of NP at the $2.9 \sigma$ level in the $B_s$-$\bar B_s$ mixing turns out to be confirmed by the present update, which includes the new D0 result on the dimuon charge asymmetry but not the new CDF measurement of $\phi_s$, being the likelihood not yet released.
}

\FullConference{35th International Conference of High Energy Physics - ICHEP2010,\\
		July 22-28, 2010\\
		Paris France}

\begin{document}

We present an update of the Unitarity Triangle Analysis (UTA) performed by the UTfit collaboration following the method described in refs.~\cite{Ciuchini:2000de,Bona:2005vz}.
Within the Standard Model (SM), we have included in $\varepsilon_K$ the contributions of $\xi$ and $\phi_\varepsilon \neq \pi/4$ which, as pointed out in~\cite{Buras:2008nn}, decrease the SM prediction for $\varepsilon_K$ by $\sim 8$\%.
We have also included the long-distance contribution calculated more recently in~\cite{Buras:2010pza}, which softens the 8\% reduction to 6\%.
In a new paper~\cite{Brod:2010mj} the perturbative calculation of the NNLO QCD corrections to the box diagram involving a top and a charm quark has been computed. This contribution, which is found to increase the theoretical prediction of $\varepsilon_K$ by 3\%, is not yet included in the UTA.

\begin{figure}[tb]	
    \center{\includegraphics[width=0.4\textwidth]{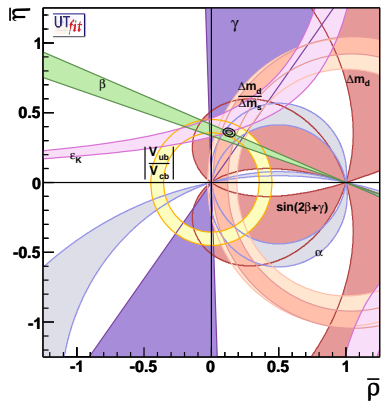}} 
    \caption{Results of the UTA within the SM. The contours display the selected 68\% and 95\% probability regions in the $(\bar \rho, \bar \eta)$-plane. The 95\% probability regions selected by the single constraints are also shown.}
\label{fig:all}
\end{figure}

We observe, as main result of the UTA, that the CKM matrix turns out to be consistently overconstraint
and the CKM parameters $\bar \rho$ and $\bar \eta$ are accurately determined: $\bar \rho=0.132\pm0.020$, $\bar \eta=0.358\pm0.012$~\cite{UTfitwebpage}.
The UTA has thus established that the CKM matrix is the dominant source of flavour mixing and CP-violation and that New Physics (NP) effects can at most represent a small correction to this picture.
We note, however, that the new contributions in $\varepsilon_K$ generate some tension in particular between the constraints provided by the experimental measurements of $\varepsilon_K$ and $\sin 2 \beta$ (see fig.~\ref{fig:all}).
As a consequence, the indirect determination of $\sin 2 \beta$ turns out to be larger than the experimental value by $\sim 2.6 \sigma$.
% \footnote{For an alternative indirect determination of $\sin 2 \beta$ which does not rely and is thus free from the hadronic uncertainty in $|V_{ub}|$, see ref.~\cite{Lunghi:2008aa}.}
We observe that the updated lattice average of the bag-parameter $B_K$~\cite{Lubicz:2010nx} further enhances this $\varepsilon_K$-$\sin 2 \beta$ tension.
This is due to the fact that new unquenched results, though compatible with older quenched results, tend to lie below them.

Recently, we have shown~\cite{Bona:2009cj} how to use the UTA to improve the prediction of BR$(B \to \tau \nu)$ in the SM, thanks to a better determination of $|V_{ub}|$ and $f_B$. Within the SM the UTA prediction for BR$(B \to \tau \nu)$ is found to deviate from the experimental measurement~\cite{HFAG} by $\sim 3.2 \sigma$. Even allowing for minimal flavour violating  NP effects, a $\sim 3.0 \sigma$ deviation from the experimental value is found.
Moreover, it is interesting to note that a large value of $|V_{ub}|$ (which is closer to some inclusive determinations) would reduce this deviation but it would enhance the tension in $\sin 2 \beta$.

We now present the update of the NP UTA, that is the UTA generalized to include possible NP effects.
% In $\varepsilon_K$ we have taken into account the effect of $\phi_{\varepsilon} \neq \pi/4$, while the $\xi$ contribution, which beyond minimal flavour violation (MFV)~\cite{D'Ambrosio:2002ex,Buras:2000dm} is affected  by a large uncertainty~\cite{Buras:2009pj}, is not included. 
This analysis consists first in generalizing the relations among the experimental observables and the elements of the CKM matrix, introducing effective model-independent parameters that quantify the deviation of the experimental results from the SM expectations.
The possible NP effects considered in the analysis are those entering neutral meson mixing.
Thanks to recent experimental developments, in fact, these $\Delta F=2$ processes turn out to provide stringent constraints on possible NP contributions.
In the case of $B_{d,s}$-$\bar B_{d,s}$ mixing, a complex effective parameter is introduced, defined as
\begin{equation}
C_{B_{d,s}}\,e^{2 i \phi_{B_{d,s}}} = \frac{\langle B_{d,s} | H_{eff}^{full}| \bar B_{d,s} \rangle}{\langle B_{d,s} | H_{eff}^{SM}| \bar B_{d,s} \rangle}\,,
\end{equation}
being $H_{eff}^{SM}$ the SM $\Delta F=2$ effective Hamiltonian and $H_{eff}^{full}$ its extension in a general NP model, and with $C_{B_{d,s}}=1$ and $\phi_{B_{d,s}}=0$ within the SM.
All the mixing observables are then expressed as a function of these parameters and the SM ones (see refs.~\cite{Bona:2005eu,Bona:2006sa,Bona:2007vi} for details).
In a similar way, for the  $K$-$\bar K$ system, one can write
\begin{equation}
\qquad C_{\epsilon_K} = \frac{Im[\langle K | H_{eff}^{full}| \bar K \rangle]}{Im[\langle K | H_{eff}^{SM}| \bar K \rangle]}\,,\qquad \qquad
C_{\Delta m_K} = \frac{Re[\langle K | H_{eff}^{full}| \bar K \rangle]}{Re[\langle K | H_{eff}^{SM}| \bar K \rangle]}\,,
\end{equation}
with $C_{\epsilon_K}=C_{\Delta m_K}=1$ within the SM.

In this way, the combined fit of all the experimental observables selects a region of the $(\bar \rho, \bar \eta)$ plane ($\bar \rho=0.135\pm0.040$, $\bar \eta=0.374\pm0.026$) which is consistent with the results of the SM analysis,
and it also constraints the effective NP parameters.

For $K$-$\bar K$ mixing, the NP parameters are found in agreement with the SM expectations. In the $B_d$ system, the mixing phase $\phi_{B_d}$ is found $\simeq 1.8 \sigma$ away from the SM expectation, reflecting the tension in $\sin 2 \beta$ discussed above.

The $B_s$-meson sector, where the tiny SM mixing phase $\sin 2 \beta_s \simeq 0.041(4)$ could be highly sensitive to a NP contribution, represents a privileged environment to search for NP.
In this sector, an important experimental progress has been achieved at the Tevatron collider since 2008 when both the CDF~\cite{Aaltonen:2007he} and D0~\cite{:2008fj} collaborations published the two-dimensional likelihood ratio for the width difference $\Delta \Gamma_s$ and the phase $\phi_s=2(\beta_s-\phi_{B_s})$, from the tagged time-dependent angular analysis of the decay $B_s \to J_{\psi} \phi$. 

In 2009 the update of the UTfit analysis of ref.~\cite{Bona:2008jn}, combining the CDF and D0 results including the D0 two-dimensional likelihood without assumptions on the strong phases, yielded $\phi_{B_s}=(-69\pm7)^\circ \cup (-19\pm8)^\circ$, which is $2.9\sigma$ away from the SM expectation $\phi_{B_s}=0$.
% A deviation of more than $2 \sigma$ was found also by the Heavy Flavour Averaging Group (HFAG)~\cite{HFAG} ($2.2 \sigma$) and by CKMfitter~\cite{Deschamps:2008de} ($2.5 \sigma$), by combining the Tevatron results with some differences in the statistical approach.

In 2010 two surprising news have arrived from the CDF and D0 experiments.
On the one hand the new CDF measurement~\cite{CDF} based on an enlarged (5.2 fb$^{-1}$) data sample  has provided a reduced significance of the deviation, from $1.8\sigma$ to $\simeq 1\sigma$.
On the other hand D0 has performed a new measurement of the dimuon charge asymmetry $a_{\mu \mu}$~\cite{Abazov:2010hj}, which points to a large value of $\phi_{B_s}$, but also to a value for the width difference $\Delta \Gamma_s$ that is significantly larger than the SM prediction.
If confirmed, the latter result would lead to one of the two (unlikely) explanations: either a huge NP contribution shows up in the tree-level observable $\Delta \Gamma_s$ or the operator product expansion badly fails (while it turns out to work well in describing b-hadron lifetimes where the same theoretical approach is adopted for the diagonal matrix element $\Gamma_{11}$ instead of $\Gamma_{12}$).

\begin{figure}[tb]	
    \center{\includegraphics[width=0.4\textwidth]{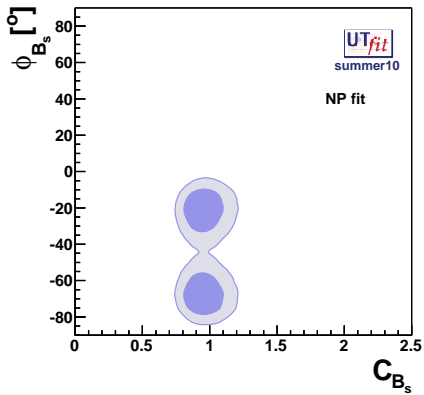}} 
    \caption{68\% (dark) and 95\% (light) probability regions in the ($C_{B_s},\phi_{B_s}$)-plane.}
\label{fig:phiBs}
\end{figure}

It will be interesting to see if these CDF and D0 results will be confirmed once the Tevatron measurements will improve.
At the moment the new CDF result is not included in the analysis since the $\Delta \Gamma_s$-$\phi_s$ likelihood has not been released yet.
Including the D0 result for $a_{\mu \mu}$, we find $\phi_{B_s}=(-68\pm8)^\circ \cup (-20\pm8)^\circ$, which confirms the $2.9\sigma$ deviation from the SM expectation $\phi_{B_s}=0$ (see fig.~\ref{fig:phiBs}).


\begin{thebibliography}{99}
%\cite{Ciuchini:2000de}
\bibitem{Ciuchini:2000de}
  M.~Ciuchini {\it et al.},
  %``2000 CKM-triangle analysis: A critical review with updated experimental
  %inputs and theoretical parameters,''
  JHEP {\bf 0107} (2001) 013
  [hep-ph/0012308].
  %%CITATION = JHEPA,0107,013;%%

%\cite{Bona:2005vz}
\bibitem{Bona:2005vz}
  M.~Bona {\it et al.}  [UTfit Collaboration],
  %``The 2004 UTfit Collaboration report on the status of the unitarity
  %triangle in the standard model,''
  JHEP {\bf 0507} (2005) 028
  [hep-ph/0501199].
  %%CITATION = JHEPA,0507,028;%%

%\cite{Buras:2008nn}
\bibitem{Buras:2008nn}
  A.~J.~Buras and D.~Guadagnoli,
  %``Correlations among new CP violating effects in Delta F = 2 observables,''
  Phys.\ Rev.\  D {\bf 78} (2008) 033005
  [0805.3887 [hep-ph]].
  %%CITATION = PHRVA,D78,033005;%%

%\cite{Buras:2010pza}
\bibitem{Buras:2010pza}
  A.~J.~Buras, D.~Guadagnoli and G.~Isidori,
  %``On epsilon_K beyond lowest order in the Operator Product Expansion,''
  Phys.\ Lett.\  B {\bf 688} (2010) 309
  [arXiv:1002.3612 [hep-ph]].
  %%CITATION = PHLTA,B688,309;%%

%\cite{Brod:2010mj}
\bibitem{Brod:2010mj}
  J.~Brod and M.~Gorbahn,
  %``epsilon_K at NNLO: The Charm-Top-Quark Contribution,''
  arXiv:1007.0684 [hep-ph].
  %%CITATION = ARXIV:1007.0684;%%

\bibitem{UTfitwebpage}
The UTfit Collaboration, http://www.utfit.org/. 

% %\cite{Lunghi:2008aa}
% \bibitem{Lunghi:2008aa}
%   E.~Lunghi and A.~Soni,
%   %``Possible Indications of New Physics in $B_d$ -mixing and in sin(2 beta)
%   %Determinations,''
%   Phys.\ Lett.\  B {\bf 666} (2008) 162
%   [arXiv:0803.4340 [hep-ph]].
%   %%CITATION = PHLTA,B666,162;%%

%\cite{Lubicz:2010nx}
\bibitem{Lubicz:2010nx}
  V.~Lubicz,
  %``Kaon physics from lattice QCD,''
  PoS {\bf LAT2009} (2009) 013
  [arXiv:1004.3473 [hep-lat]].
  %%CITATION = POSCI,LAT2009,013;%%


%\cite{Bona:2009cj}
\bibitem{Bona:2009cj}
  M.~Bona {\it et al.}  [UTfit Collaboration],
  %``An Improved Standard Model Prediction Of BR(B -> tau nu) And Its
  %Implications For New Physics,''
  Phys.\ Lett.\  B {\bf 687} (2010) 61
  [arXiv:0908.3470 [hep-ph]].
  %%CITATION = PHLTA,B687,61;%%


\bibitem{HFAG}
The Heavy Flavour Averaging Group (HFAG), http://www.slac.stanford.edu/xorg/hfag/.\\
G.~De Nardo on behalf of the BaBar Collaboration, these proceedings.

% %\cite{D'Ambrosio:2002ex}
% \bibitem{D'Ambrosio:2002ex}
%   G.~D'Ambrosio {\it et al.},
%   %``Minimal flavour violation: An effective field theory approach,''
%   Nucl.\ Phys.\  B {\bf 645} (2002) 155
%   [hep-ph/0207036].
%   %%CITATION = NUPHA,B645,155;%%
% 
% %\cite{Buras:2000dm}
% \bibitem{Buras:2000dm}
%   A.~J.~Buras {\it et al.},
%   %``Universal unitarity triangle and physics beyond the standard model,''
%   Phys.\ Lett.\  B {\bf 500} (2001) 161
%   [hep-ph/0007085].
%   %%CITATION = PHLTA,B500,161;%%
% 
% %\cite{Buras:2009pj}
% \bibitem{Buras:2009pj}
%   A.~J.~Buras and D.~Guadagnoli,
%   %``On the consistency between the observed amount of CP violation in the
%   %$K^{-}$ and Bd-systems within minimal flavor violation,''
%   Phys.\ Rev.\  D {\bf 79} (2009) 053010
%   [arXiv:0901.2056 [hep-ph]].
%   %%CITATION = PHRVA,D79,053010;%%


%\cite{Bona:2005eu}
\bibitem{Bona:2005eu}
  M.~Bona {\it et al.}  [UTfit Collaboration],
  %``The UTfit collaboration report on the status of the unitarity triangle
  %beyond the standard model. I: Model-independent analysis and minimal  flavour
  %violation,''
  JHEP {\bf 0603} (2006) 080
  [hep-ph/0509219].
  %%CITATION = JHEPA,0603,080;%%

%\cite{Bona:2006sa}
\bibitem{Bona:2006sa}
  M.~Bona {\it et al.}  [UTfit Collaboration],
  %``The UTfit collaboration report on the unitarity triangle beyond the
  %standard model: Spring 2006,''
  Phys.\ Rev.\ Lett.\  {\bf 97} (2006) 151803
  [hep-ph/0605213].
  %%CITATION = PRLTA,97,151803;%%

%\cite{Bona:2007vi}
\bibitem{Bona:2007vi}
  M.~Bona {\it et al.}  [UTfit Collaboration],
  %``Model-independent constraints on Delta F=2 operators and the scale of New
  %Physics,''
  JHEP {\bf 0803} (2008) 049
  [0707.0636 [hep-ph]].
  %%CITATION = JHEPA,0803,049;%%

%\cite{Aaltonen:2007he}
\bibitem{Aaltonen:2007he}
  T.~Aaltonen {\it et al.}  [CDF Collaboration],
  %``First Flavor-Tagged Determination of Bounds on Mixing-Induced CP Violation
  %in Bs -> J/psi phi Decays,''
  Phys.\ Rev.\ Lett.\  {\bf 100} (2008) 161802
  [0712.2397 [hep-ex]].
  %%CITATION = PRLTA,100,161802;%%

%\cite{:2008fj}
\bibitem{:2008fj}
  V.~M.~Abazov {\it et al.}  [D0 Collaboration],
  %``Measurement of $\boldmath {B_s^0}$ mixing parameters from the flavor-tagged
  %decay $B^0_s \to J/\psi \phi$,''
  Phys.\ Rev.\ Lett.\  {\bf 101} (2008) 241801
  [0802.2255 [hep-ex]].
  %%CITATION = PRLTA,101,241801;%%

%\cite{Bona:2008jn}
\bibitem{Bona:2008jn}
  M.~Bona {\it et al.}  [UTfit Collaboration],
  %``First Evidence of New Physics in $b \longleftrightarrow s$ Transitions,''
  PMC Phys.\  A {\bf 3} (2009) 6
  [arXiv:0803.0659 [hep-ph]].
  %%CITATION = PMCPA,A3,6;%%

% %\cite{Deschamps:2008de}
% \bibitem{Deschamps:2008de}
%   O.~Deschamps,
%   %``CKM global fit and constraints on New Physics in the $B$ meson mixing,''
%   arXiv:0810.3139 [hep-ph].
%   %%CITATION = ARXIV:0810.3139;%%

\bibitem{CDF}
G.~Giurgiu on behalf of the CDF Collaboration, these proceedings. 

%\cite{Abazov:2010hj}
\bibitem{Abazov:2010hj}
  V.~M.~Abazov {\it et al.}  [D0 Collaboration],
  %``Evidence for an anomalous like-sign dimuon charge asymmetry,''
  Phys.\ Rev.\ Lett.\  {\bf 105} (2010) 081801
  [arXiv:1007.0395 [hep-ex]].
  %%CITATION = PRLTA,105,081801;%%



\end{thebibliography}
\end{document}